\documentclass{article}

    \usepackage[preprint]{neurips_2020}

\usepackage[utf8]{inputenc} 
\usepackage[T1]{fontenc}    
\usepackage{hyperref}       
\usepackage{url}            
\usepackage{booktabs}       
\usepackage{amsfonts}       
\usepackage{nicefrac}       
\usepackage{microtype}      
\usepackage{algorithm}
\usepackage{graphicx}
\usepackage{subfig}
\usepackage{siunitx}

\title{Convolution Neural Networks for diagnosing colon and lung cancer histopathological images}

\author{
  Sanidhya Mangal \\
  Engineerbabu \\
  \texttt{sanidhya.mangal@engineerbabu.in}\\
  \And
  Aanchal Chaurasia\\
  Engineerbabu \\
  \texttt{aanchal@engineerbabu.com}\\
  \And
  Ayush Khajanchi \\
  Emorphis Technologies \\
  \texttt{ayush.khajanchi@icorprated.com}\\
}

\begin{document}

\maketitle

\begin{abstract}
 Lung and Colon cancer are one of the leading causes of mortality and morbidity in adults.
 Histopathological diagnosis is one of the key components to discern cancer type.
 The aim of the present research is to propose a computer aided diagnosis system for diagnosing
 squamous cell carcinomas and adenocarcinomas of lung as well as adenocarcinomas of colon using convolutional neural networks
 by evaluating the digital pathology images for these cancers.
 Hereby, rendering artificial intelligence as useful technology in the near future.
 A total of 2500 digital images were acquired from LC25000 dataset containing 5000 images for each class.
 A shallow neural network architecture was used classify the histopathological slides into
 squamous cell carcinomas, adenocarcinomas and benign for the lung.
 Similar model was used to classify adenocarcinomas and benign for colon.
 The diagnostic accuracy of more than $\SI{97}{\percent}$ and $\SI{96}{\percent}$ was recorded for lung and colon respectively.
\end{abstract}

\section{Introduction}
According to the World health organization(WHO) cancer is the leading cause of morality in the world. 
Lung cancer is the most commonly diagnosed cancer ($\SI{11.6}{\percent}$ of the total cases) and the leading cause of cancer death 
($\SI{18.4}{\percent}$ of the total cancer deaths), while colorectal cancer contributes to ($\SI{9.2}{\percent}$) for mortality ~\citeauthor{bray2018global}.
A rising trend has been reported around the globe for the incidence of malignant tumors which may be attributed to an increase in population. 
Malignancy can occur to any age group by histopathological type is diagnosed mostly in elderly age group of 50-60 years ~\citeauthor{arslan2018analysis}. 
It has been predicted that cancer mortality might bump up to $\SI{60}{\percent}$ untill 2035 ~\citeauthor{araghi2019global}.

Lung cells become cancerous when they mutate to grow uncontrollably and form a cluster called a tumor.~\citeauthor{AmericanCancerSocitey}
The worldwide increase in lung cancer has been attributed to different factors but mainly to the exposure to breathing dangerous or toxic substances
and the rise in the number of aged people in the society. The symptoms, however, are not likely to be observed until it has spread to the other parts of the body,
which makes it harder to treat it. 

Although lung cancer can occur in people who have never smoked, it usually is the greatest risk for people who do.
Adenocarcinoma and squamous cell carcinoma are the most commonly occurring types of lung cancer while the other histological types include small as well as large cell carcinomas.
Adenocarcinoma of lung cancer usually occurs in current or former smokers but is also prevalent in non-smokers. This is more prone to occur in women and youth and 
is found in the outer parts of the lungs even before it spreads.
Squamous cell carcinomas are also associated with the history of smoking. Small and Large cell carcinoma, on the other hand, 
can develop in any part of the lung and has the tendency to grow and unroll rapidly making it harder to medicate.

Colon, the final part of our digestive system, when host cancerous cells, could cause colon cancer. Colon cancer is not age-specific but it typically affects older adults.
It usually begins as small, noncancerous (benign) clumps of cells called polyps that form on the inner side of the colon.
Over time, some of these polyps can develop to cause colon cancers.

In most colon cancers, a tumor is formed when healthy cells in the lining of the colon or rectum grow uncontrollably.
Adenocarcinomas of the colon or rectum develop in the lining of the large intestine occuring in the epithelial cells and then spread to the other layers.
Mucinous adenocarcinomas and signet ring cell adenocarcinoma are two less common subtypes of adenocarcinoma but are aggressive and difficult to medicate.
The changes can occur over the years in one’s body which is reliant on factors such as gender, ethnicity, age, smoking patterns, and socio-economic conditions.
However, if a person has some unusual inherited syndrome, the changes can transpire in a small duration of some months.

The aim of the research paper revolves around measuring the potency of the proposed algorithm of the Convolutional Neural Network (CNN) ~\citeauthor{lecun1995convolutional} to detect the common types of lung and colon cancer in the human body. 
The architecture of the algorithm keeps in mind the patterns of the neurons and their connectivity inside the human brain which also attributes for it’s low pre-processing required in comparison to other classification algorithms. 
The ability of the algorithm to learn characteristics overpowers the primitive method wherein filters are hand-engineered.
The ConvNet takes input in the form of images attributes weight (learnable weights and biases) to multiple features in the image,
and is able to differentiate one image from the other. We use Histopathology slides~\citeauthor{gurcan2009histopathological} as a dataset, 
since the preparation process preserves the underlying tissue architecture it provides an eclectic view of disease and its effect on tissues.
Additionally,histopathology image renders as a ‘gold standard’ in diagnosing almost distinct cancer types~\citeauthor{rubin2008rubin}.

In the later section the paper is organized in the following manner: Section~\ref{sec:relatedwork} provides an insight about previously explored research in the current domain.
Section~\ref{sec:dataset} provided a brief introduction about LC25000 dataset.
Section~\ref{sec:deeplearningcnn} covers a short introduction on CNN,
In addition to this Section~\ref{sec:cnnarchitecture} elaborates the cnn architecture used in training both the models.
Section~\ref{sec:results} reports all the experimental results and findings.
Finally, Section~\ref{sec:conclusion} concludes our experiment while presenting some insights on future work.

\section{Related Work}
\label{sec:relatedwork}
~\citeauthor{doi2007computer} explored the potential for automatic image processing around 4 decades ago but it is still challenging due to complexity of images to analyze.
Back then feature extraction was a key step in adopting machine learning based computer-aided diagnosis (CAD). Different ontologies of cancer has been investigated in,
~\citeauthor{te2000automatic},~\citeauthor{beller2005example},~\citeauthor{yin1994computerized},~\citeauthor{aerts2014decoding},~\citeauthor{eltonsy2007concentric},
~\citeauthor{wei2005computer},~\citeauthor{hawkins2014predicting},~\citeauthor{barata2012system},~\citeauthor{barata2014bag},~\citeauthor{han2015texture},
~\citeauthor{sadeghi2013detection},~\citeauthor{zikic2012decision},~\citeauthor{meier2013hybrid}.
Moreover, ~\citeauthor{munir2019cancer} provides a detailed overview on cancer diagnosis by conducting experiments on several deep learning techniques.
It also provides a comparison of various predominant architectures for each technique.
 
Also, ~\citeauthor{coudray2018classification} trained an inceptionv3 ~\citeauthor{szegedy2016rethinking} model for classification and mutation
from non–small cell lung images of adenocarcinomas and squamous cell carcinomas achieving a mean area under curve(auc) of 0.97.
They also mutated the result for ten most common genes for lung adenocarcinomas. Similarly ~\citeauthor{ardila2019end} predicts the risk of lung cancer using deep learning techniques
by computing prior and current tomography of the patients.
~\citeauthor{lakshmanaprabu2019optimal} explored the CT scans using deep neural networks and linear discriminant analysis for automated diagnosis.
~\citeauthor{sirinukunwattana2016locality} used spatially constrained CNN to perform nuclei detection and classification of cancerous tissues in colon histology images.
 
Recently, ~\citeauthor{abbas2020histopathological} presented a comparative study on histopathology diagnosis on squamous cell carcinomas using CNNs.
It compares various CNN architectures such as AlexNet, VGG-16, ResNet achieving an F-1 score of 0.97. Similarly, ~\citeauthor{bukhari2020histological}
presents a comparative analysis on colonic adenocarcinomas using variations of ResNet architecture achieving a baseline accuracy of $\SI{93}{\percent}$.

\section{LC25000 Dataset}
\label{sec:dataset}
\begin{figure}[!h]
  \centering
  \subfloat[lung adenocarcinomas]{
  \includegraphics[width=4.2cm, height=3cm]{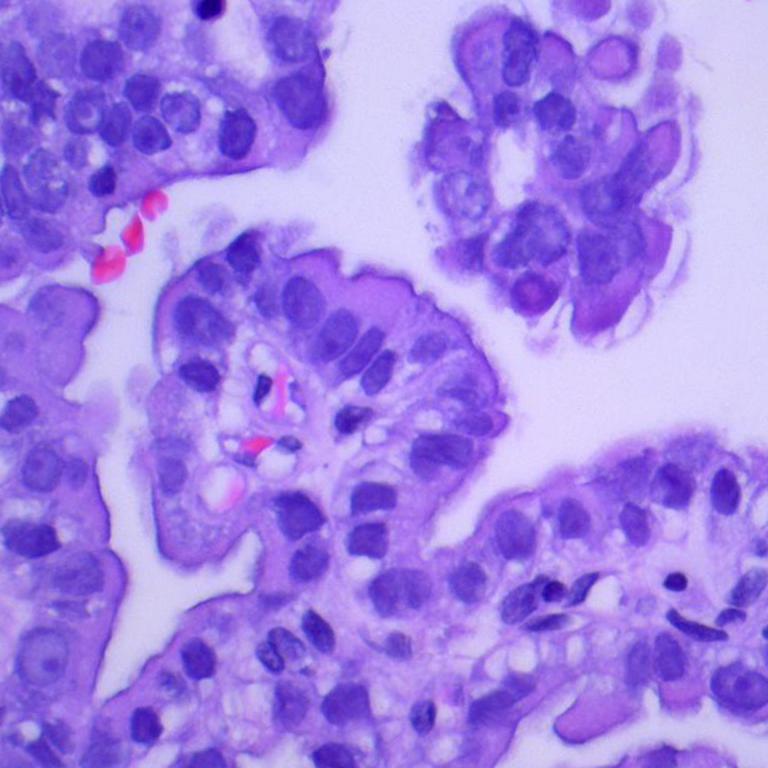}
  }
  \subfloat[lung squamous cell carcinomas]{
  \includegraphics[width=4.2cm, height=3cm]{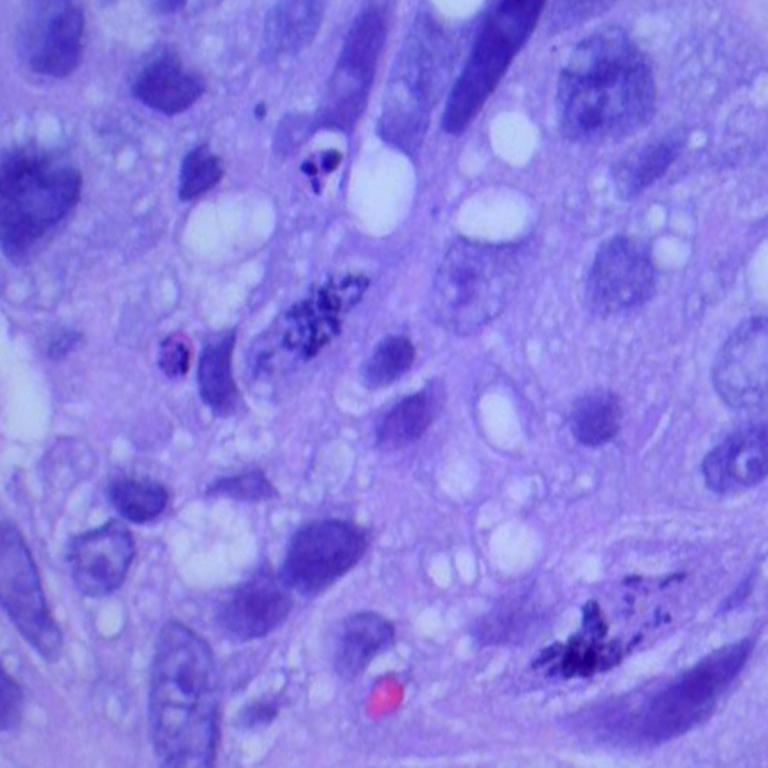}
  }
  \subfloat[lung benign]{
  \includegraphics[width=4.2cm, height=3cm]{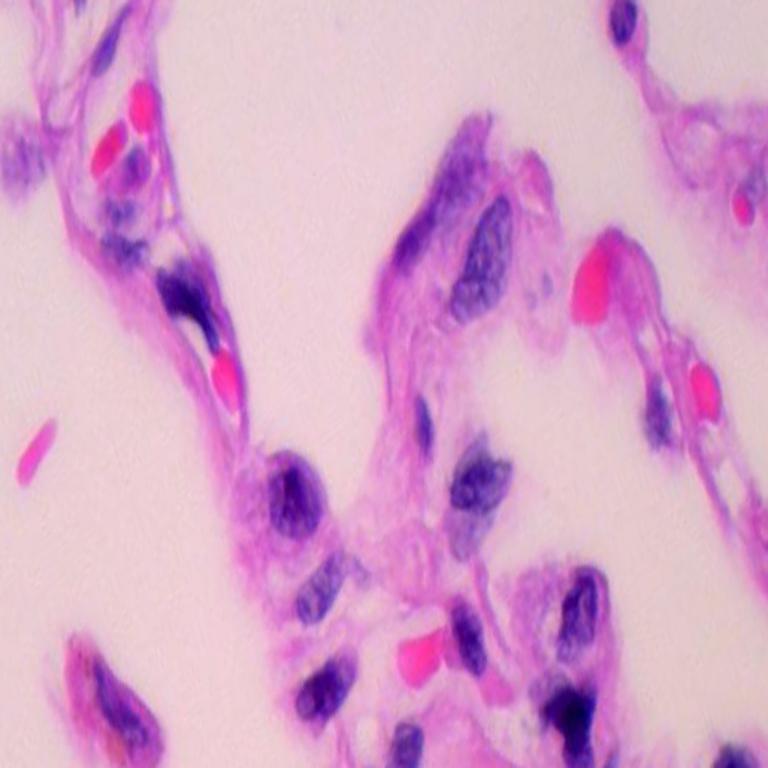}
  } 
  \hspace{0mm}
  \subfloat[colon adenocarcinomas]{
  \includegraphics[width=4.2cm, height=3cm]{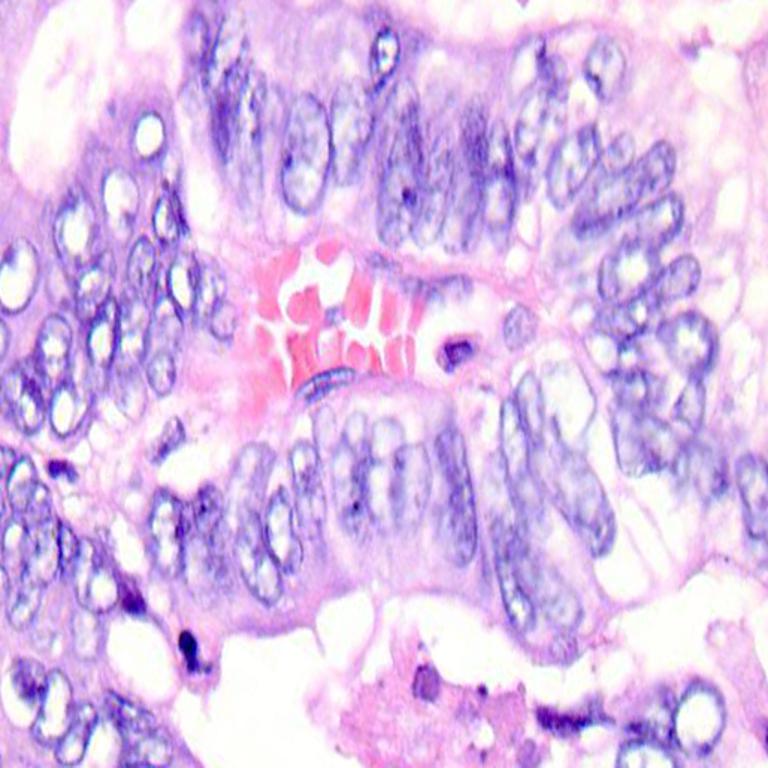}
  }
  \subfloat[colon benign]{
  \includegraphics[width=4.2cm, height=3cm]{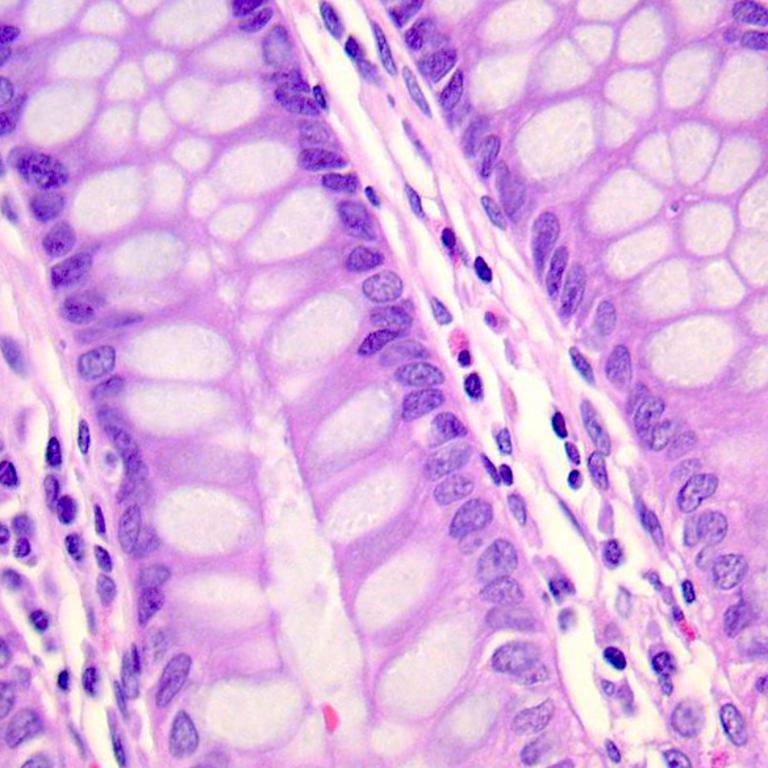}
  }
  \caption{(a) and (b) is an example image of adenocarcinomas and squamous cell carcinomas cancer types for the lung, 
  (c) represents the benign histopathology of lung. 
  Similarly in (d) adenocarcinomas cancer for colon is described where in (e) illustrates benign class for colon
  }
  \label{fig:datagrid}
\end{figure}

A brief introduction on dataset is provided followed by all the data preprocessing steps. 
The LC25000 Dataset ~\citeauthor{borkowski2019lung} contains microsopic images of lung and colon. 
The dataset can be bifuragted into five different classes namely, 
\textbf{lung adenocarcinomas, lung squamous cell carcinomas, lung benign, colon adenocarcinomas and colon benign} 
each containing \verb+5000+ images. 
Figure~\ref{fig:datagrid} describes some sample images corresponding to above mentioned classes. 
Original dataset contains only \verb+750+ images lung and \verb+500+ images of colon with pixel size of \verb+1024x768+,
later it was converted into square of \verb+768x768+ pixels. 
Augmentor was used to expand the dataset into \verb+25000+ images with the help of rotation and flips. 

Before feeding augmented data, it undergo some preprocessing. 
Initially data was sampled into \verb+4500+ and \verb+500+ datapoints for training and test set respectively
for each class using random sampling proposed in ~\citeauthor{practiceofstats}. 
Furthermore images were resized to \verb+150x150+ pixels,
along with some randomized shear, zoom transformation followd by normaliztion of images.


\section{Deep Learning Approach Using CNN}
\label{sec:deeplearningcnn}
Image classification is a challenging task for the visual content, 
particularly microscopic images for example histopathological images 
due to high convolution of inter-intraclass dependencies. 
The underlying structures are complex and interwoven due to similar structural morphological textures. 
Figure~\ref{fig:datagrid} presents some of the complex textures present in histopathology of images. 
Deep learning is prevalent due to its ability to learn features directly from the input, 
providing us a window to avoid arduous feature extraction processes ~\citeauthor{bengio2013representation}. 
One of the key features of deep learning is to discover abstract level features and 
then deep dive for extracting structural semantics in the feature map. 
In recent years, deep learning, especially CNNs has proven to be an effective tool for 
classifying and diagnosing medical images 
~\citeauthor{shen2017deep}, ~\citeauthor{kermany2018identifying}, ~\citeauthor{lee2017deep}, ~\citeauthor{suzuki2017overview}
In nutshell, CNN contains multiple trainable layers which could be stacked together along with a supervised classifier 
to learn feature maps from the given input data feed. Input data feed could be either digital or signal data 
such as audio, video, images and time-series. 
For example, upon considering a coloured image it is a feature map of 3D tensor, 
i.e., a 2D tensor for each colour channel.

\begin{figure}[!h]
  \centering
  \includegraphics[width=10cm, height=4cm]{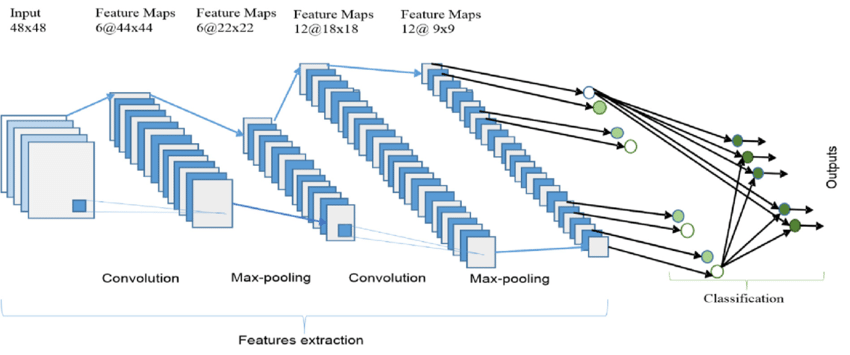}
  \caption{Describes one of the most commonly used cnn architectures, ~\citeauthor{articlecnnarch}}
  \label{fig:cnnarch}
\end{figure}

CNN architectures are composed of mainly three layers which are: 
convolution layer, max pooling layer and fully connected layer or dense layer. 
These layers could be stacked in multiple combinations to produce a CNN. 
An example of typical CNN is show in Figure~\ref{fig:cnnarch}.

In a typical CNN, convolution layer acts as a key component for any given architecture. 
Convolution layers compute a dot product between weights and input signal connected to that local region. 
The set of weights which are convoluted along the input vector is called \verb+kernel or filters+. 
Each filter is small but extends across the full depth of the input volume. 
For image inputs a typical size of filter is generally (3x3, 5x5, 8x8). 
These weights are shared across neurons so that filters can learn all the geometrical structures from the image. 
The distance between applications of these filters is termed as \verb+stride+. 
If hyperparameter stride is smaller than filter size than overlapping convolutions are applied to the image.

It is a common practice to insert a pooling in between two convolution layers in order to downsample the image 
along volume component, This is crucial to reduce progressively the spatial size of the representation. 
Thus, reducing the number of parameters and computations required by the network helps in the overfitting control. 
The pooling operation resizes the images along height and width discarding activation. 
In practice, \verb+max pooling+ operation which provides a window of selecting the maximum value from input patch among neighborhoods 
has shown better results ~\citeauthor{scherer2010evaluation}.

In fully connected layer or \verb+dense layer+, full connection is established between activations of input and their activation. 
Computation is done with the help of matrix multiplication along with successive bias offset. 
The last fully connected layer contains the final output such as probability density or \verb+logit+ values 
~\citeauthor{spanhol2016breast}, ~\citeauthor{krizhevsky2012imagenet}.


\section{CNN Architecture and Training Strategy}
\label{sec:cnnarchitecture}
In order to classify the (citeauthor dataset name) dataset we constructed the deep CNN with following layers and parameters:

\paragraph{Input Layer}This layer is used to load data and feed it to the first convolution layer, 
In our case the input is an image of size 150x150 pixels with colour channels which is 3 for RGB. 

\paragraph{Convolution Layer}This layer is used to convolve the input image with trainable filters to learn the geo spatial structure of images, 
this model contains three convolution layers with filter size 3x3, stride set to 2 and padding kept the same. 
First layer contains 32 filters, followed by two layers with 64 filters each and they are initialized with Gaussian distribution. 
In addition to this, ReLU activation is applied for non linear operation to improve the performance ~\citeauthor{behnke2003hierarchical}

\paragraph{Pooling Layer}Pooling operation is used for downsampling the output images received from the convolution layer. 
There is one pooling layer after each convolution layer with pooling size of 2, padding set to valid. 
All the pooling layers use the most common max pooling operation.

\paragraph{Flatten Layer}This layer is used to convert the output from the convolution layer into a 1D tensor to connect a dense layer 
or fully connected layer.

\paragraph{Fully connected layer or dense layer}These layers treat the input as a simple vector and produce an output in a form of vector. 
Two dense layers are used in this model, first one contains 512 neurons and last one contains 3 and 2 neurons for 
lung and colon cancer respectively depending on the input class. 
Output from the last fully connected layer could be activated with the help of softmax activation.
\begin{equation}
  \label{eq:softmax}
  Softmax(x_i) = \frac{e^{x_i}}{\sum_j e^{x_j}}
\end{equation}

\paragraph{Dropout Layer} In order to prevent overfitting of the model layers we use a dropout layer in between fully connected 
layers which randomly drops neurons from both visible and hidden layers ~\citeauthor{srivastava2014dropout}. 

Table~\ref{tab:cnnlayers} Illustrates the parameters of the layers, where CONV+POOL stands for convolution layer followed by a pooling layer 
and FC by fully connected layer or dense layer.

\begin{table}[!h]
  \caption{Summary of CNN Layers}
  \label{tab:cnnlayers}
  \centering
  \begin{tabular}{lccccc}
    \toprule
    & \multicolumn{5}{c}{Layers}                   \\
    \cmidrule(r){2-6}
    &1   & 2   & 3 & 4 & 5\\
    \midrule
    Type & CONV+POOL  & CONV+POOL &   CONV+POOL & FC &   FC  \\
    Channels     & 32 &   64  &   64   & 512 &   3 or 2  \\
    Filter Size     & 3x3  &   3x3   &   3x3 & - & -   \\
    Convolution Strides     & 2x2  &   2x2   &   2x2 & - & -  \\
    Pooling Size     & 2x2  &   2x2   &   2x2 & - & -   \\
    Pooling Strides & 1x1  &   1x1   &   1x1  & - & -  \\
    \bottomrule
  \end{tabular}
\end{table}

For all the CNN modes a similar training protocol was used, purely supervised in nature. 
The RMSprop method proposed by ~\citeauthor{rmsprop} with backpropagation was used to compute gradient and 
a mini batch size of 32 was used to update network weights, 
with starting learning rate of 10-4, in conjunction with $\rho = 0.9$ and $\epsilon = 1e-7$. 
Categorical cross entropy loss ~\ref{eq:xent} is used to ensure that performance of the model is maintained throughout the training process.
The CNN was trained for 100 iterations.
\begin{equation}
  \label{eq:xent}
  E_{\mbox{\tiny entropy}} = -\sum_n^N \sum_k^c
   t_k^n \ln y_k^n
\end{equation}

\section{Results}
\label{sec:results}
To ensure that classifiers generalize well, 
the data was split into three categories, with 80-10-10 of data into training, 
testing and validation set respectively into distinct sets. This protocol was applied independently for both lung and colon cancer images. 
When discussing medical image processing it can be evaluated in two ways, first one is at patient level, i.e., 
determining the amount of images classified correctly for each patient. 
Secondly, it can be evaluated at the image where we calculate the number of cancer images classified correctly. 
In LC25000 dataset \ref{sec:dataset} no information about patients was provided hence we decided to move forward with a later method to evaluate the model performance. 
All the CNN models were trained on Google's Colab using TensorFlow framework ~\citeauthor{tensorflow2015-whitepaper}. 
These models would be made available in h5 format at \url{https://github.com/sanidhyamangal/lung_colon_cancer_research}. 
Training each model took around 45 minutes.

\begin{figure}[!h]
  \centering
  \subfloat[colon filter 1]{
  \includegraphics[width=4.2cm, height=3cm]{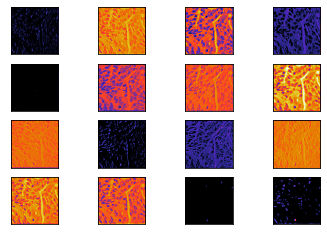}
  }
  \subfloat[colon filter 2]{
  \includegraphics[width=4.2cm, height=3cm]{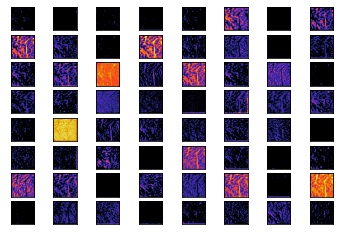}
  }
  \subfloat[colon filter 3]{
  \includegraphics[width=4.2cm, height=3cm]{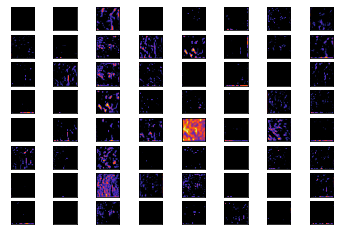}
  }
  \hspace{0mm}
  \subfloat[lung filter 1]{
  \includegraphics[width=4.2cm, height=3cm]{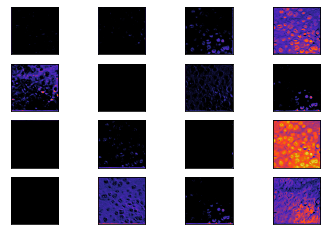}
  }
  \subfloat[lung filter 2]{
  \includegraphics[width=4.2cm, height=3cm]{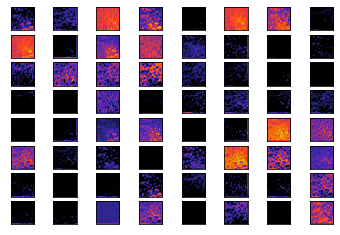}
  }
  \subfloat[lung filter 3]{
  \includegraphics[width=4.2cm, height=3cm]{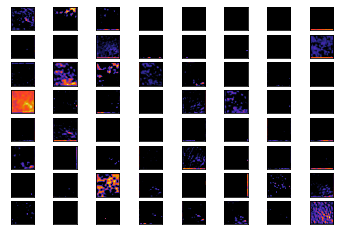}
  }
  \caption{
    Feature maps learned by convolution network layers, (a), (d) describes filters form first convolution layer for colon and lung models respectively, 
    Similarly, (b), (e), (c), (f) illustrates the filters from second and third convolution layers respectively.
    }
  \label{fig:convfilters}
\end{figure}

Deep Learning techniques are one of the advanced machine learning techniques which do not require the design of 
feature extraction by domain experts but model learns by itself. 
We can learn the feature detectors learned by models, 
considering the weights learned by feature maps. 
Figure~\ref{fig:convfilters} describes feature maps learned by all the convolution layers for both the models. 
We can visualize the filters at image level and also filters resemble like Gabor filters(
~\citeauthor{fogel1989gabor}, ~\citeauthor{bovik1990multichannel}, ~\citeauthor{Zeiler_2014}).

\begin{figure}[!h]
  \centering
  \subfloat[Colon Loss]{
  \includegraphics[width=6cm, height=3.6cm]{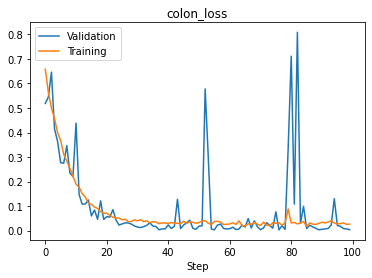}
  }
  \subfloat[Colon Accuracy]{
  \includegraphics[width=6cm, height=3.6cm]{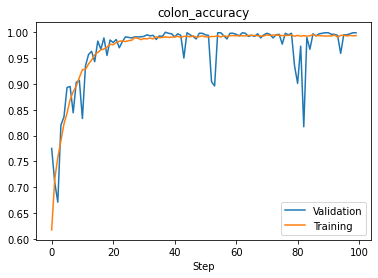}
  }
  \hspace{0mm}
  \subfloat[Lung Loss]{
  \includegraphics[width=6cm, height=3.6cm]{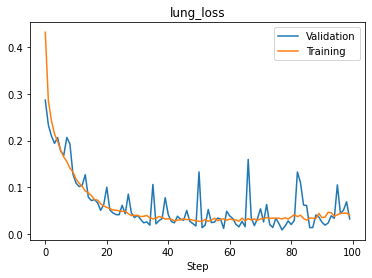}
  }
  \subfloat[Lung Accuracy]{
  \includegraphics[width=6cm, height=3.6cm]{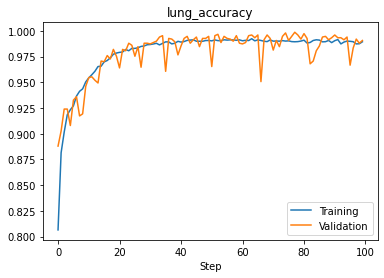}
  }
  \caption{Delineates the accuracy and loss metrics plot over steps. (a), (b) reports the loss and accuracy metrics over all the 100 steps for the colon model.
  Similarly, loss and accuracy for lung model is described in (c), (d)}
  \label{fig:lossaccuracygraph}
\end{figure}

To better assess performance metrics, Figure~\ref{fig:lossaccuracygraph} visualizes the plot between epochs vs loss and accuracy.
It is quite transparent that there are jitters in all the accuracy and loss graphs due to the dropout layer which helps the neural network generalize.
However, there is a slight aberration in the plots of colon, i.e., validation loss increases initially upto 5 epochs then starts to converge.
Also, it can be inferred that there are large spikes in both accuracy and loss for the colon model at around $58^{th}$ and $80^{th}$ epoch.
For both the models it could also be extrapolated that both the models took around 20 epochs to converge.

Table~\ref{tab:modelperformance} reports the performace metrics for both the models, in both the modes, i.e.,
training and validation at image level.

\begin{table}[h]
  \caption{Performance metrics of all the models}
  \label{tab:modelperformance}
  \centering
  \begin{tabular}{lccc}
    \toprule
    Type & Rule & Accuracy  & Loss \\
    Lung & Training & $\SI{97.9216 \pm 0.0266}{\percent}$ & $\SI{05.8323 \pm 0.06149}{\percent}$\\
    & Validation & $\SI{97.8987 \pm 0.0234}{\percent}$& $\SI{06.1145 \pm 0.05719}{\percent}$ \\
    \midrule
    Colon & Training & $\SI{96.9503 \pm 0.0614}{\percent}$ & $\SI{07.9340 \pm 0.1200}{\percent}$ \\
     & Validation & $\SI{96.611 \pm 0.0614}{\percent}$ & $\SI{09.7141 \pm 0.1681}{\percent}$ \\
    \bottomrule
  \end{tabular}
\end{table}

\section{Conclusions}
\label{sec:conclusion}
In this paper, we have presented a set of experiments conducted on the LC25000 dataset using a deep learning approach.
We have shown that we could use a shallow CNN architecture, that has been designed for classifying color images of objects,
and adapt it to the classification of lung and colon histopathological images.
We have also proposed a training and evaluation strategy for training the CNN architecture, it allows to deal with the high-resolution of
these textured images without converting those images to low-resolution images. Our experimental results obtained on the LC25000 showed improved accuracy obtained by CNN
when compared to traditional machine learning models and deep convolutional neural networks models leveraging transfer learning trained on the same dataset
but with state of the art texture descriptors.
Future work can explore different CNN architectures and the optimization of the hyperparameters.
Also, strategies to apply neural style transfer for generating interclass images for different histopathology.
In addition to this generative models could be used to generate histopathological images for visualizing and exploring mutations on different ontologies.

\begin{ack}
We would like to acknowledge Mayank Pratap Singh, Aditi Chaurasia, Sumit Yadav, and Prachi Bundela for helpful discussions.
Ravinder Singh shared his script for train test split with us and much-needed support with \LaTeX\ typesetting.
We would like to thank the developers of TensorFlow.
We would also like to thank Engineerbabu IT Services PVT LTD.
for providing computational resources.

\end{ack}

\medskip

\small

\bibliography{lung_colon}
\bibliographystyle{rusnat}
\end{document}